# Numerical Study on Mechanism of Small Vortex Generation in Boundary Layer Transition

Ping Lu[1] and Chaoqun Liu[2]

UNIVERSITY OF TEXAS AT ARLINGTON, ALRINGTON, TX 76019, USA
CLIU@UTA.EDU

**The small vortex generation is a key issue of the mechanism for late flow transition and turbulence generation. It was widely accepted that small length vortices were generated by large vortex breakdown. According to our recent DNS, we find that the hairpin vortex structure is very stable and never breaks down to small pieces. On the other hand, we recognize that there are strong positive spikes besides the ring neck in the spanwise direction. The strongly positive spikes are caused by second sweeps which are generated by perfectly circular and perpendicularly standing vortex rings. The second sweep brings energy from the invisid region downdraft to the bottom of the boundary layers, which generates high shear layers around the positive spikes. Since the high shear layer is not stable, all small length scales (turbulence) are generated around high shear layers especially near the wall surface (bottom of boundary layers). This happens near the ring neck in the streamwise direction and besides the original vortex legs in the spanwise direction. The small length scales then rise up from the wall surface and convect to the upper boundary layer. Therefore, the small length scales (turbulence) are not generated by "large vortex breakdown", but by "high shear layers" produced by the positive spikes and the wall surface.**

## Nomenclature

$M_\infty$ = Mach number
$\delta_{in}$ = inflow displacement thickness
$T_\infty$ = free stream temperature
$Lz_{out}$ = height at outflow boundary
$Lx$ = length of computational domain along x direction
$Ly$ = length of computational domain along y direction
$x_{in}$ = distance between leading edge of flat plate and upstream boundary of computational domain
$A_{2d}$ = amplitude of 2D inlet disturbance
$\omega$ = frequency of inlet disturbance
$\alpha_{2d}, \alpha_{3d}$ = two and three dimensional streamwise wave number of inlet disturbance
$\beta$ = spanwise wave number of inlet disturbance
$\gamma$ = ratio of specific heats

$Re$ = Reynolds number
$T_w$ = wall temperature
$Lz_{in}$ = height at inflow boundary

$A_{3d}$ = amplitude of 3D inlet disturbance

$R$ = ideal gas constant
$\mu_\infty$ = viscosity

## I. Introduction

THE transition process from laminar to turbulent flow in boundary layers is a basic scientific problem in modern fluid mechanics and has been the subject of study for over a century. After over a century study on the turbulence, the linear and early weakly non-linear stages of flow transition are well understood. However, for the late non-linear transition stages, there are many questions still remaining for research (Kleiser et al, 1991; Borodulin et al, 2002；Bake et al 2002, Rist et al, 2002; Kachanov, 2003; Guo et al, 2004; 2010; Meyer, Kloker, and Rist U., 2003). Ring-like vortices play a key role in flow transition. They generate rapid downward jets (second sweeps) which induce positive spikes and bring energetic freestream fluid into the boundary layer, working together with upward jets (ejection) to mix the boundary layer. In other words, the ring-like vortex formation and development are the key feature for late flow transition in boundary layer. It appears that there is no turbulence without ring-like vortices. Anyway, many questions related to late flow transition have not been answered or have been misinterpreted. In order to achieve a deep understanding on the late flow transition in a boundary layer, we recently conducted a high-resolution, high-order DNS with 1920x241x128 gird points and about 600,000 time steps to study the mechanisms of the late stages of flow transition in a boundary layer. Many new findings were made and new mechanisms were revealed (Chen et al 2010a, 2010b, 2010c; Liu et al, 2010a, 2010b, 2010c, 2010d). Here, we use the $\lambda_2$ criterion (Jeong & Hussain, 1995) for visualization.

---

[1] PhD Student, AIAA Member, University of Texas at Arlington, USA
[2] Professor & AIAA Associate Fellow, University of Texas at Arlington, USA

1
American Institute of Aeronautics and Astronautics

The concept of "large vortex breakdown to turbulence" was widely accepted (Singer et al 1994, Schlichting and Gersten,2000) for flow transition as the mechanism of turbulence generation. However, according to our new DNS, we found that the phenomena of "vortex breaks down to smaller structures" are theoretically impossible and never happen in practice. A ring is the only type of vortex which can exist inside a fluid field and a vortex cannot breakdown to smaller structures according to the Helmholtz vorticity conservation law.

This question is easily raised that where the small length scales (turbulence) come from? First we will look back at the concept of high-shear (HS) layer which was first introduced by Kovasznay, L.S., Komoda, H. and Vasudeva, B.R., in 1962. At the end of 90s., the relation between the sweep events and the high shear layers has been investigated. Originally, people thought that the vortex rings were generated by HS which was located around legs. However, later they found that the location of the vortex ring generation was near the vortex ring neck but not near the vortex legs. The position of ring-like vortex generation was not located at the same place where the large shear stress appeared. They did not coincide (Bake et al 2002). Later it was commonly thought the ring-like formation process was similar to the Crow theory (Crow S C, 1970; Borodulin et al, 2002). However, this hypothesis has not been supported with strong evidences. Recently, by careful study on our new DNS, we confirm that the small length scales are indeed produced by high shears. Of course the HS is not located around vortex legs as people thought while ago, but it is around the positive spikes which are generated by second sweeps besides the ring-like vortex legs near the wall surface.

## II. Case setup and DNS validation

### 2.1 Case setup

The computational domain is displayed in Figure 1. The grid level is 1920×128×241, representing the number of grids in streamwise (*x*), spanwise (*y*), and wall normal (*z*) directions. The grid is stretched in the normal direction and uniform in the streamwise and spanwise directions. The length of the first grid interval in the normal direction at the entrance is found to be 0.43 in wall units ($Y^+$=0.43).

The parallel computation is accomplished through the Message Passing Interface (MPI) together with domain decomposition in the streamwise direction Figure 2. The computational domain is partitioned into N equally-sized sub-domains along the streamwise direction. N is the number of processors used in the parallel computation. The flow parameters, including Mach number, Reynolds number etc are listed in Table 1. Here, $x_{in}$ represents the distance between leading edge and inlet, $Lx$, $Ly$, $Lz_{in}$ are the lengths of the computational domain in x-, y-, and z-directions, respectively, and $T_w$ is the wall temperature.

| $M_\infty$ | $Re$ | $x_{in}$ | $Lx$ | $Ly$ | $Lz_{in}$ | $T_w$ | $T_\infty$ |
|---|---|---|---|---|---|---|---|
| 0.5 | 1000 | 300.79 $\delta_{in}$ | 798.03 $\delta_{in}$ | 22 $\delta_{in}$ | 40 $\delta_{in}$ | 273.15K | 273.15K |

**Table 1. Flow parameters**

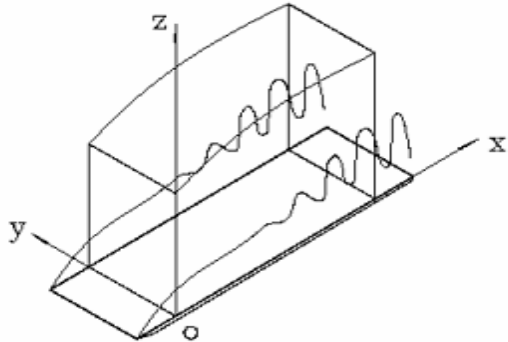

**Figure 1. Computation domain**

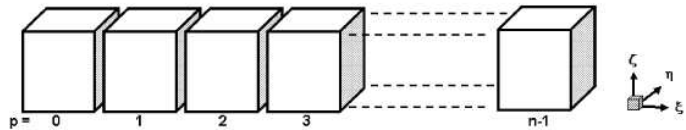

**Figure 2. Domain decomposition along streamwise direction in the computational space**

**2.2 Code validation and DNS results visualization**
In order to make sure that our DNS code and DNS results are correct and accurate, a number of verifications and validations have been conducted.

*2.2.1 Comparison with Linear Theory*
Figure 3 compares the velocity profile of the T-S wave given by our code and linear theory and Figure 4 is a comparison of the perturbation amplification rate between DNS and LST. The agreement between linear theory and our numerical results is pretty well.

*2.2.2 Grid Convergence*
The skin friction coefficient calculated from the time- and spanwise-averaged profile is displayed in Figure 5. The spatial evolution of skin friction coefficients of laminar flow is also plotted out for comparison. It is observed from these figures that the sharp growth of the skin-friction coefficient occurs after $x \approx 450\delta_{in}$, which is defined as the 'onset point'. The skin friction coefficient after transition is in good agreement with the flat-plate theory of turbulent boundary layer by Cousteix in 1989 (Ducros, 1996). Figure 6 also shows that we get grid convergence in velocity profile.

*2.2.3 Comparison with Log Law*
Time- and spanwise- averaged streamwise velocity profiles for various streamwise locations in two different grid levels are shown in Figure 6. The inflow velocity profiles at $x = 300.79\delta_{in}$ is a typical laminar flow velocity profile. At $x = 632.33\delta_{in}$, the mean velocity profile approaches a turbulent flow velocity profile (Log law). This comparison shows that we get the turbulent flow velocity profile and we get the grid convergence. Note that the velocity profile given by Singer et all (1994) has large discrepancy in velocity profile with Log law (see Figure 6(c)). On the other hand, we cannot find the Log Law in the paper given by Borodulin et al (2002)

*2.2.4 Comparison with Experiment*
A vortex identification method introduced by Jeong & Hussain (1995) is applied to visualize the vortex structures by using an iso-surface of a $\lambda_2$-eigenvalue. The vortex cores are found by the location of the inflection points of the pressure in a plane perpendicular to the vortex tube axis. The pressure inflection points surround the pressure minimum that occurs in the vicinity of the vortex core. By this $\lambda_2$-eigenvalue visualization method, the vortex structures shaped by the nonlinear evolution of T-S waves in the transition process are shown in Figure 7(a). The evolution details are briefly studied in our previous paper (Lin et al 2009) and the formation of ring-like vortices chains is consistent with the experimental work (Lee C B & Li R Q, 2007, Figure 7(b)) and previous numerical simulation by Rist and his co-authors (Bake et al 2002).

### 2.2.5 Comparison with Other DNS

Although we cannot compare our DNS results with those given by Borodulin et al (2002) quantitatively and we cannot find Log law in their DNS, we still can find the shear layer structure are very similar in two DNS computations in Figure 8. This cannot happen by accident.

All these verifications and validations show that our code is correct and our DNS results are reliable.

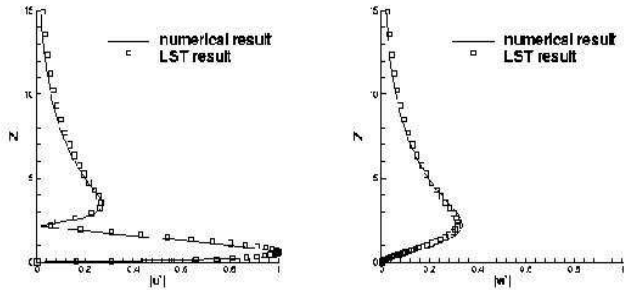

**Figure 3. Comparison of the numerical and LST velocity profiles at Rex=394300**

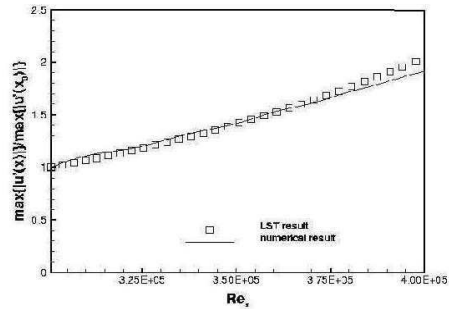

**Figure 4. Comparison of the perturbation amplification rate between DNS and LST**

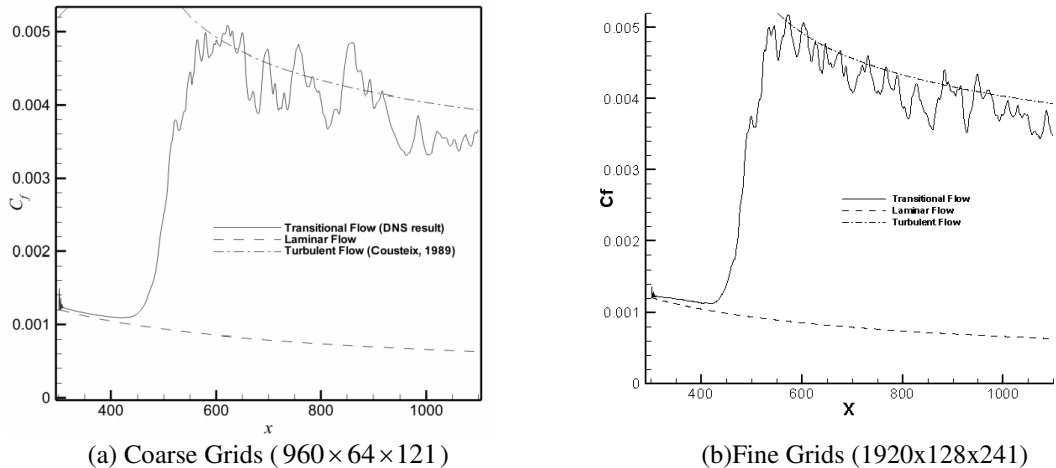

(a) Coarse Grids ($960 \times 64 \times 121$)          (b) Fine Grids (1920x128x241)
**Figure 5. Streamwise evolutions of the time-and spanwise-averaged skin-friction coefficient**

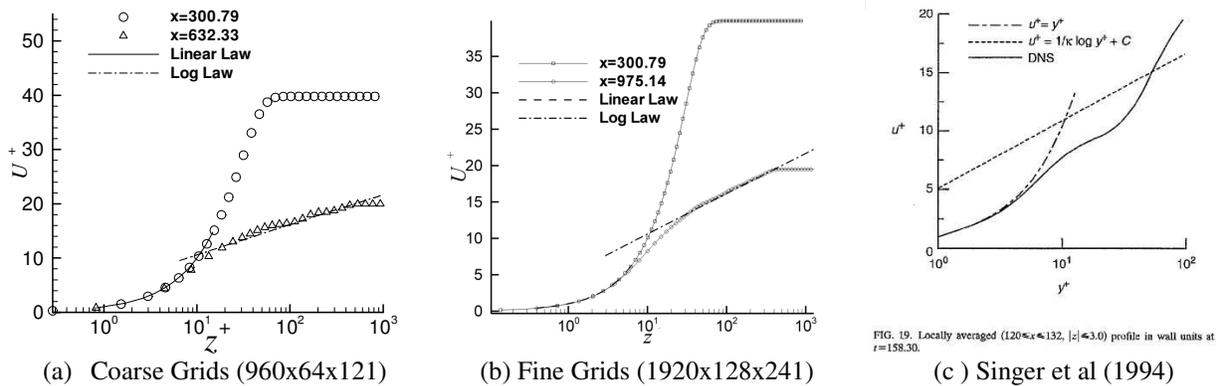

**Figure 6. Log-linear plots of the time-and spanwise-averaged velocity profile in wall unit.**

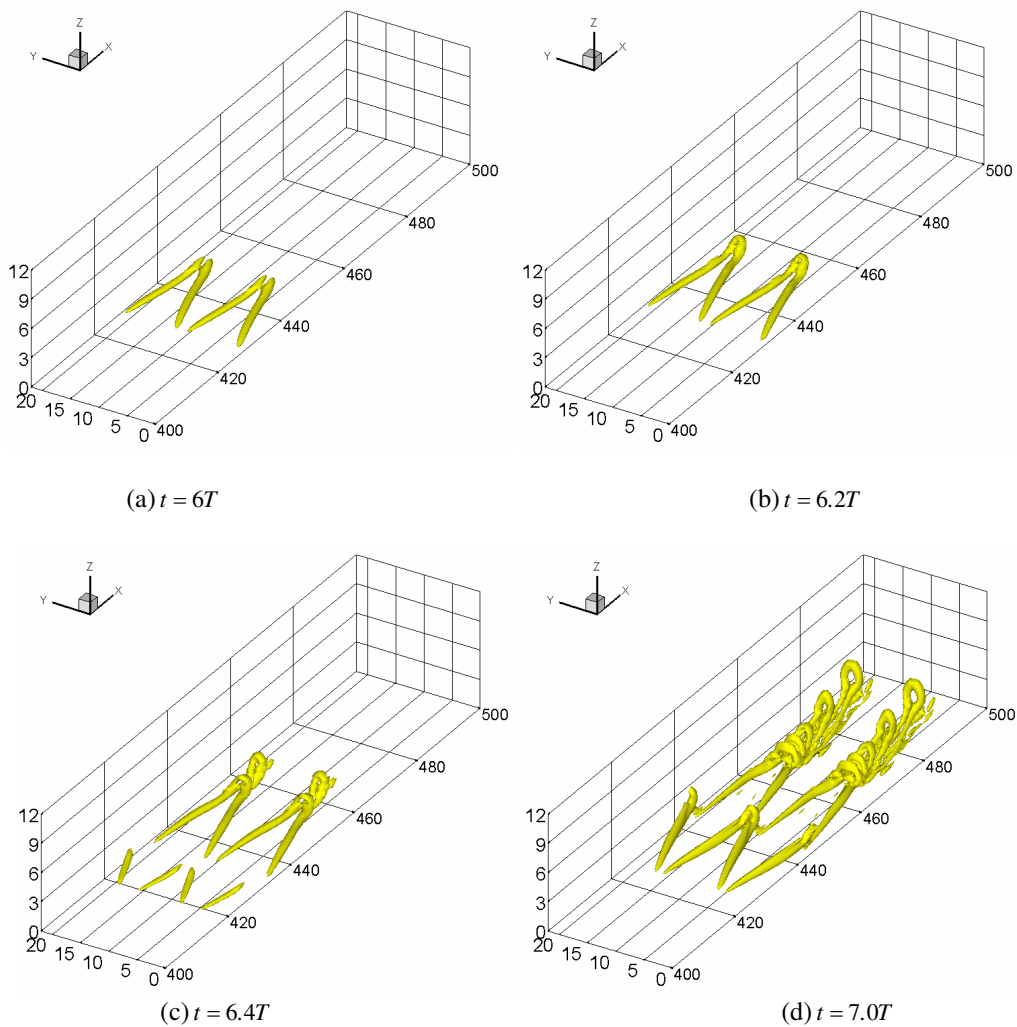

**Figure 7a. The evolution of vortex structures at the late-stage of transition
(Where $T$ is the period of T-S wave)**

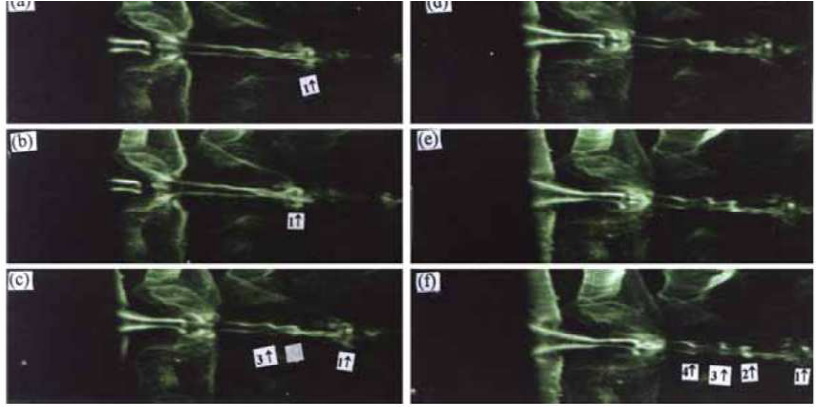

**Figure 7b. Evolution of the ring-like vortex chain by experiment (Lee et al, 2007)**

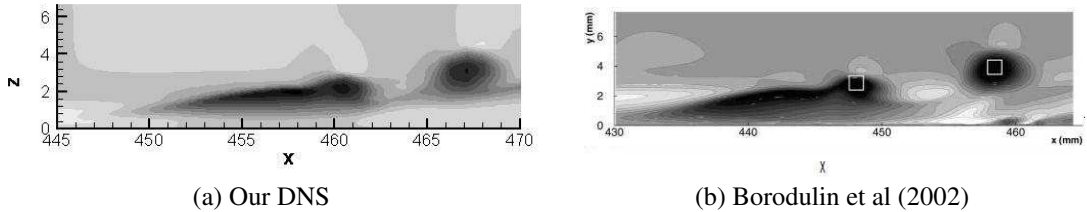

(a) Our DNS  (b) Borodulin et al (2002)

**Figure 8. Qualitatively Comparison of contours of streamwise velocity disturbance *u* in the (*x, z*)-plane (Light shades of gray correspond to high values)**

In boundary layer flows, viscosity is non-negligible. So standard approaches, such as integrating vortex lines, using minimum pressure or maximum vorticity, may lead to improper vortex identification. Jeong and Hussain have established a robust criterion for identification of vortex (or coherent) structures in viscous flows based on the eigenvalues of the symmetric $3\times 3$ tensor

$$M_{ij} := \sum_{k=1}^{3} \Omega_{ik}\Omega_{kj} + S_{ik}S_{kj} , \qquad (1)$$

where

$$\Omega_{ij} := \frac{1}{2}(\frac{\partial u_i}{\partial x_j}+\frac{\partial u_j}{\partial x_i}) \text{ and } S_{ij} := \frac{1}{2}(\frac{\partial u_i}{\partial x_j}-\frac{\partial u_j}{\partial x_i}) \qquad (2)$$

represent the symmetric and anti-symmetric components of the velocity gradient tensor, $\nabla u$. Given the three (real) eigenvalues of M at each grid point, a vortex core is identified as any contiguous region having two negative eigenvalues. If the eigenvalues are sorted such that $\lambda_1 \leq \lambda_2 \leq \lambda_3$, then any region for which $\lambda_2 < 0$ corresponds to a vortex core. One advantage of this approach is that vortices can be identified as isosurfaces of a well-defined scalar field. Moreover, the criterion $\lambda_2(x) < 0$ is scale invariant, so in principle there is no ambiguity in selecting which isosurface value to render. In practice, one usually biases the isosurface to a value that is below zero by a small fraction of the full dynamic range in order to avoid noise in regions where the velocity is close to zero. Here, we use the $\lambda_2$ criterion (Jeong & Hussain, 1995) for visualization.

## 2.3 Questions on "hairpin vortex breakdown"

It was widely accepted that the last stage of laminar flow transition is caused by "hairpin vortex breakdown". Figure 9 shows a schematic of flow transition on a flat plate in the book by Schilichting et al (2000). However, the hairpin structure is pretty stable and never breaks down according to our new DNS (Figure 10).

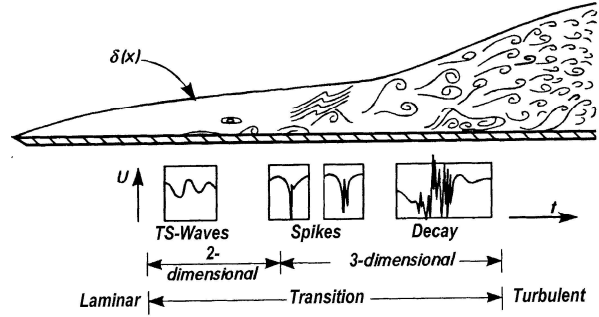

**Figure 9. Schematic of flow transition on a flat plate**
*(Copy of Figure 15.38, Page 474, Book of Boundary Layer Theory by Schilichting et al, 2000)*

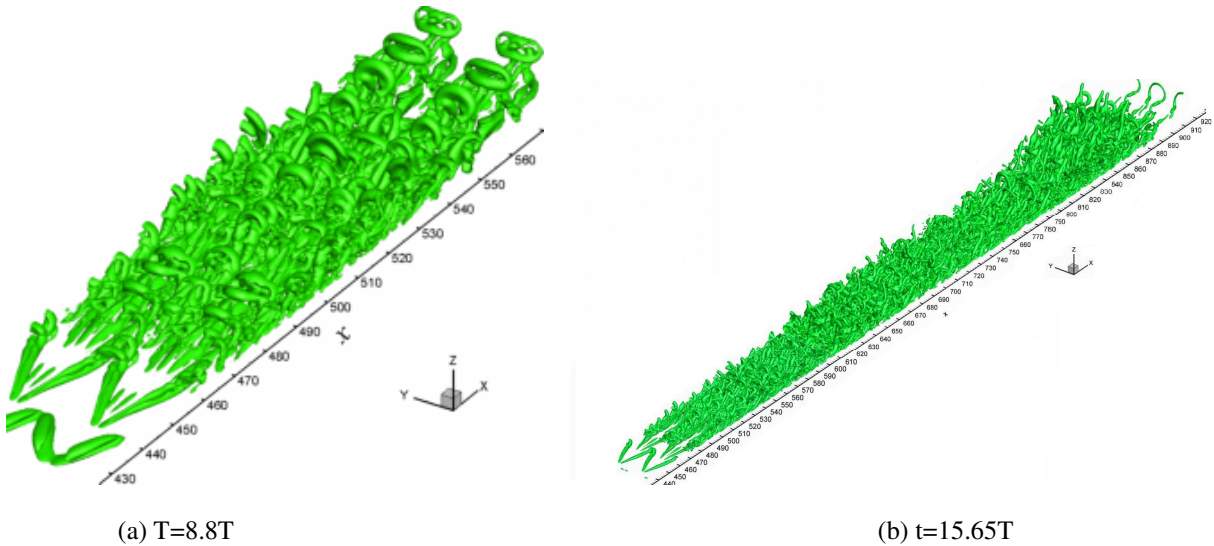

(a) T=8.8T  (b) t=15.65T

**Figure 10. Multiple ring generation and hairpin vortex structure**

## III. Mechanism of small length scale generation

The results presented in this section are taken at a streamwise range $430 < x < 530$ and instant time $t = 8.0T$ shown in figure 11. Through these plots where two slices are selected (x=508.633 and y=4.0 respectively), a general scenario of formation and development of small length scale structures at the late stages of flow transition can be found clearly. In order to fully understand the mechanism of generation of small length scale, we need to focus in more details on how and where the positive spike and HS layer are generated first.

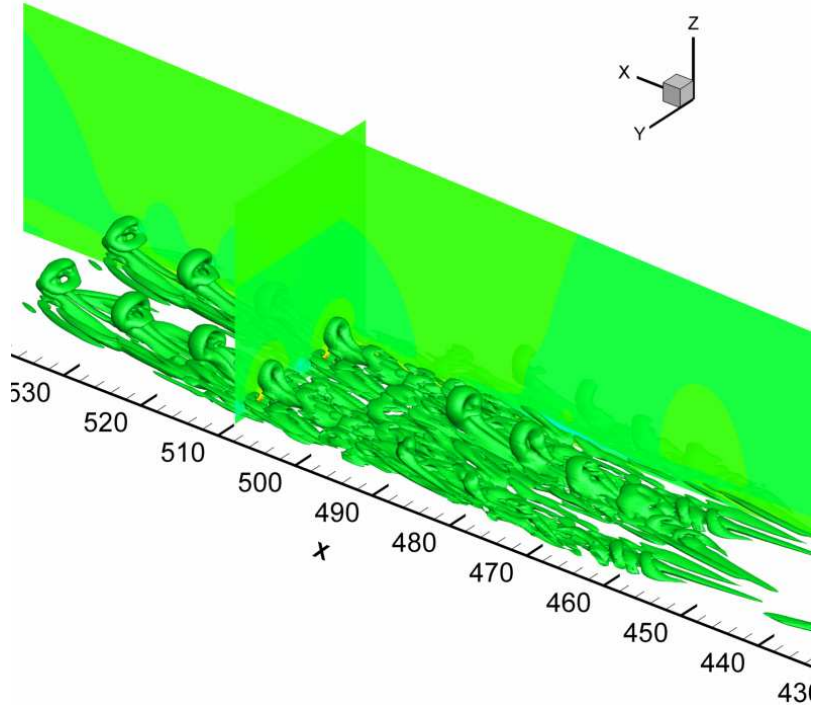

**Figure 11. Visualization of flow transition at t=8.0T based on eigenvalue $\lambda_2$ and two slices**

### 3.1 Formation of positive spike and high shear layer

Our numerical results confirm that there is a second sweep excited by every ring-like vortex (Guo et al, 2004). Combined with the first sweep generated by the original λ vortex legs, it forms a strong positive spike which generates strong downward motion to the bottom of the boundary layer. From Figure 12, we can find that there are intensively high speed streaks (red color) associated with positive spikes observed in the near wall region.

Figure 13 illustrates that the appearance of the high-shear layers around positive spikes produced by downward jets. The downward jets are induced by the ring-like vortices which must be perfectly circular and perpendicularly standing. It leads to formation of very strong second sweep and high-shear layers near the wall region at two sides of λ vortex legs. Meanwhile, at the position x=480, the updraft motion which is related to the first ejection also forms a HS-layer due to the reduction of velocity by first ejection.

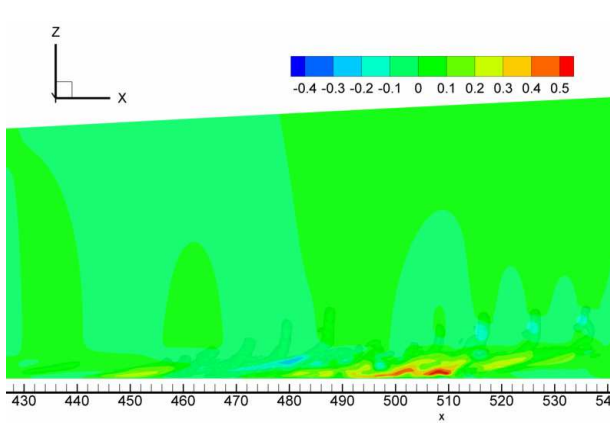

**Figure 12. Perturbation velocity contour and eigenvalue $\lambda_2$**

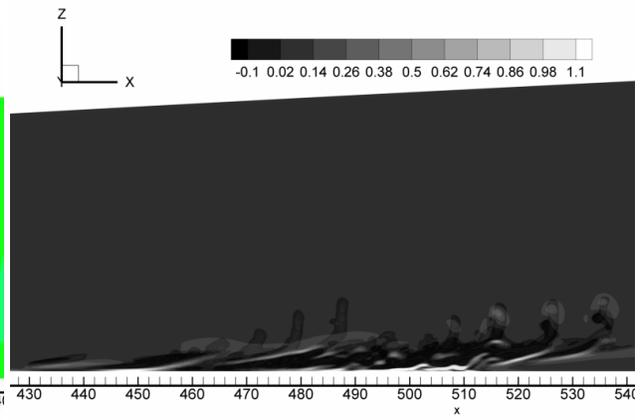

**Figure 13. Vorticity contour and eigenvalue $\lambda_2$**

## 3.2 The shape of each ring-like vortices at early stage

Figure 14 demonstrates some physical phenomenon about the shape of each ring-like vortices at the early stage (t=7.0T) of flow transition. At x=486, the leading ring-like vortex is almost perfectly circular. The reason is that the leading ring-like vortex is formed at the upper edge of the boundary layer and almost inside the inviscid area where the flow is isotropic. Meanwhile, another question will be raised up that why the leading ring-like vortex stands almost perpendicularly and some later ring-like vortices (at x= 476) do not stand perpendicularly. The reason is that both the top and bottom of the leading ring-like vortex are located in the inviscid region(U=0.99Ue) and move at the same speed, however, the later rings are not located fully in the inviscid region where the top of the ring moves faster than the bottom of the ring because of the boundary layer mean velocity profile. In addition, we can answer the question why the ring legs are inclined. This is because they are located inside the boundary layer and must be stretched and inclined due to the boundary layer mean velocity profile.

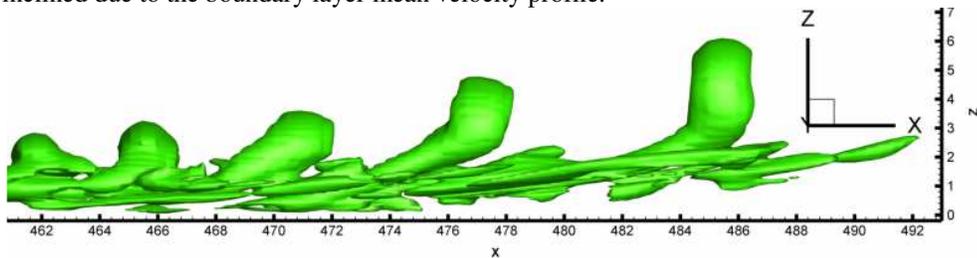

**Figure 14. Side view at t=7.0T**

## 3.3 Small length scale vortices generation near the wall

Figure 15(a)-(b) are contours at enlarged section x=508.633 and Figures 15(c)-(d) are for the enlarged section at y=4.0 with contours of the streamwise disturbance velocity and spanwise component of vorticity in the (*x,z*) and (*y,z*)-planes respectively. The second sweep movement induced by ring-like vortices working together with first sweep will lead to huge energy and momentum transformation from high energy inviscid zones to low energy zones near the bottom of the boundary layers. We can observe that **small length scales are all generated around the high speed region due to high shear layer especially between the positive spikes and solid wall surface.**

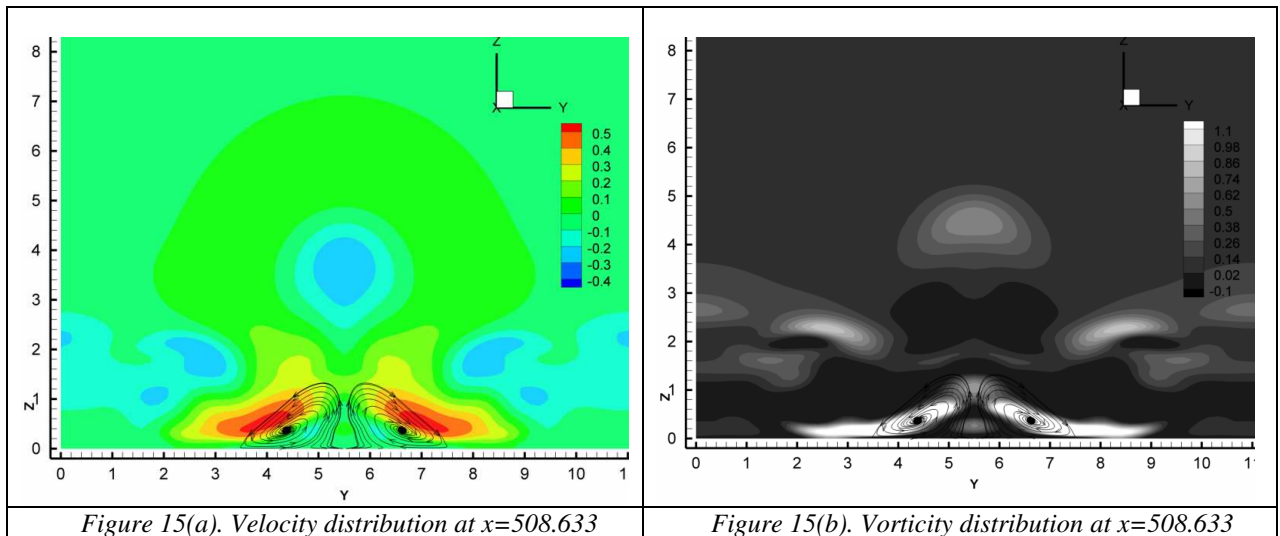

| *Figure 15(a). Velocity distribution at x=508.633* | *Figure 15(b). Vorticity distribution at x=508.633* |
|---|---|

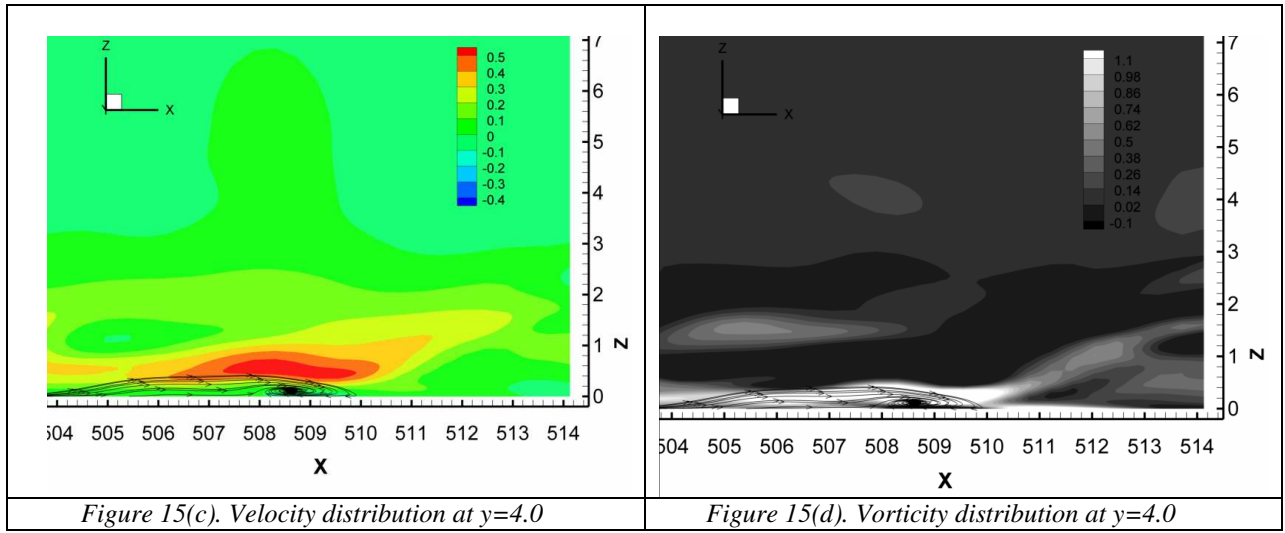

*Figure 15(c). Velocity distribution at y=4.0*  *Figure 15(d). Vorticity distribution at y=4.0*

The small vortex generation is a key mechanism for the late flow transition and turbulence generation. In order to see more details of small structures, the instantaneous snapshots are taken using $\lambda_2$ with small negative value selected for visualization. From the Figures 16(a) and (b), we can observe that high shear layer (HS) is generated by positive spike at the bottom of the boundary layer. However, since HS is not stable, it breaks down (HS breakdown but not vortex breakdown) and forms small vortices. Figures 16(c) and (d) demonstrate that the small length scales (turbulence) are generated near the wall surface in the normal direction, near the ring neck in the streamwise direction and besides the original vortex legs in the spanwise direction. Then they are stretched from the wall surface and convect to the upper boundary layer.

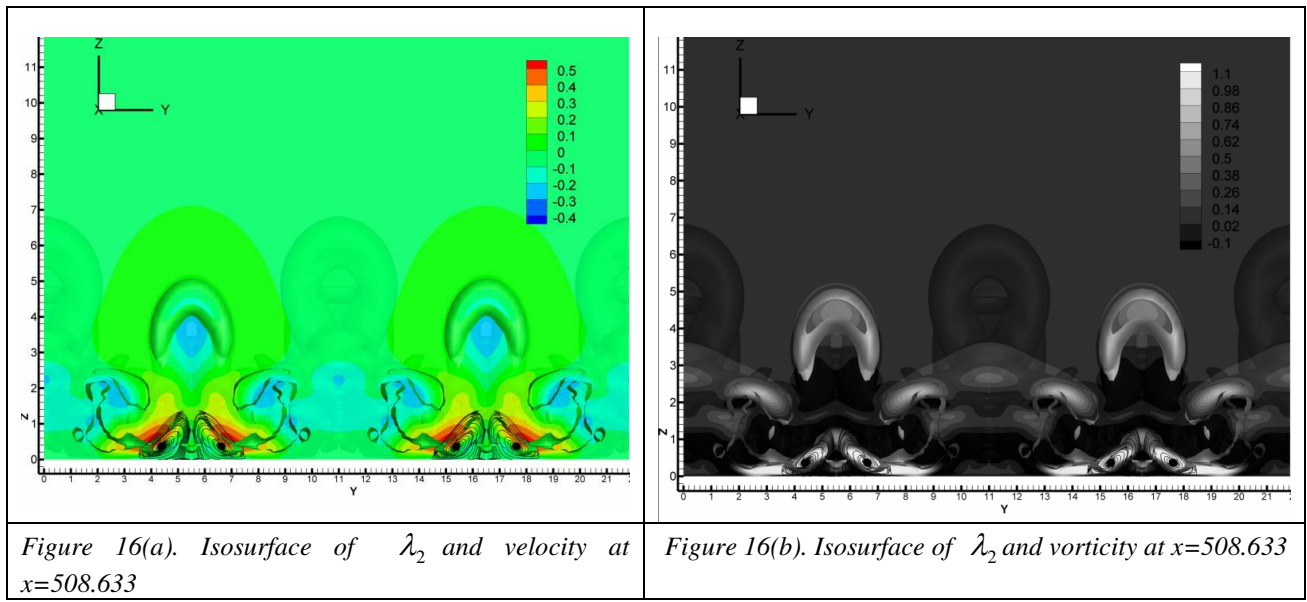

*Figure 16(a). Isosurface of $\lambda_2$ and velocity at x=508.633*  *Figure 16(b). Isosurface of $\lambda_2$ and vorticity at x=508.633*

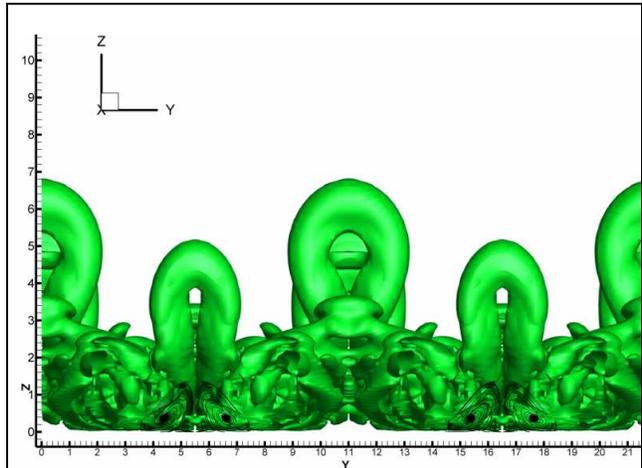

*Figure 16(c). Isosurface of $\lambda_2$ and streamtrace at x=508.633*

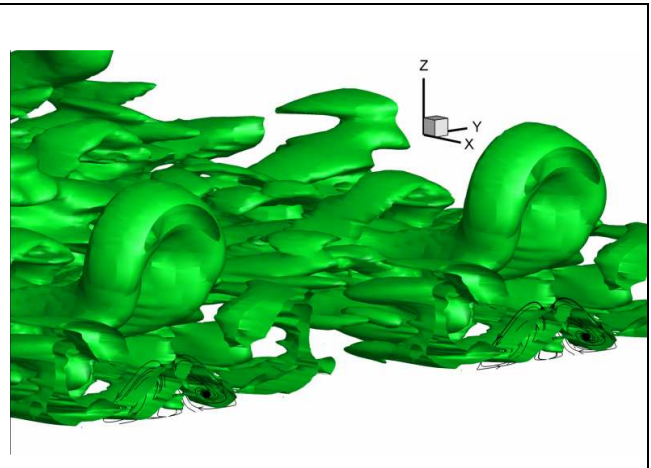

*Figure 16(d). Isosurface of $\lambda_2$ and streamtrace at x=508.633*

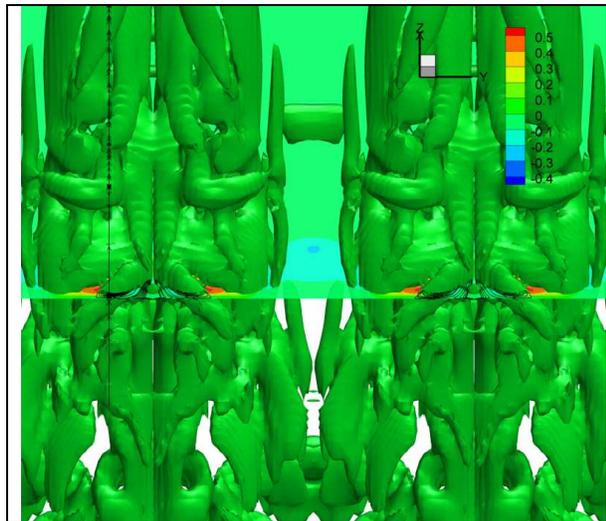

*Figure 17(a). Bottom view and velocity distribution at x=508.633*

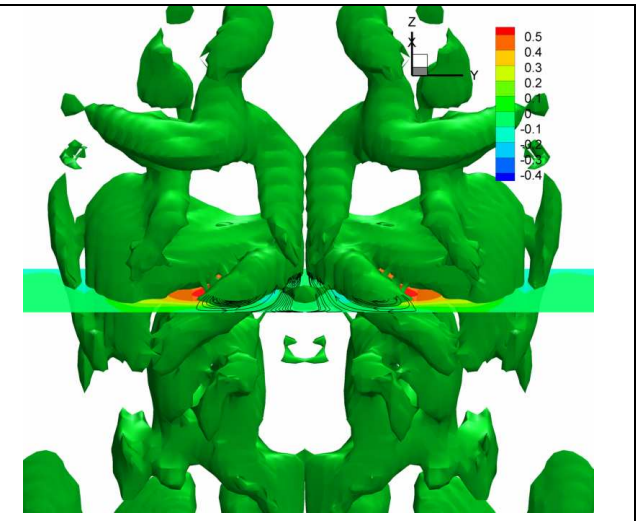

*Figure 17(b). Enlarged bottom view and velocity distribution at x=508.633*

Figure 17(a) and (b) are viewed from the plate bottom using the visualization of $\lambda_2$ and perturbation velocity. We can find that each of the small length scale has two legs which lie down on the wall surface. It proves that the vorticity can be only given or generated in boundaries, rather than generated inside the flow field. Actually, the wall surface is the sole source of the vorticity generation.

## IV. Comparison of two different time steps

In order to detect the presence of small length scales produced by high shear layer, to investigate this phenomenon quantitatively, and to understand how the strong downdraft jets disappears when ring-like vortices move downstream, we look at the time steps t=8.075T and t=8.175T. From Figure 18, we can find that the legs of small vortices at t=8.075T are much stretched towards to the center at x=510.01 in comparison with one at the position x=508.633.

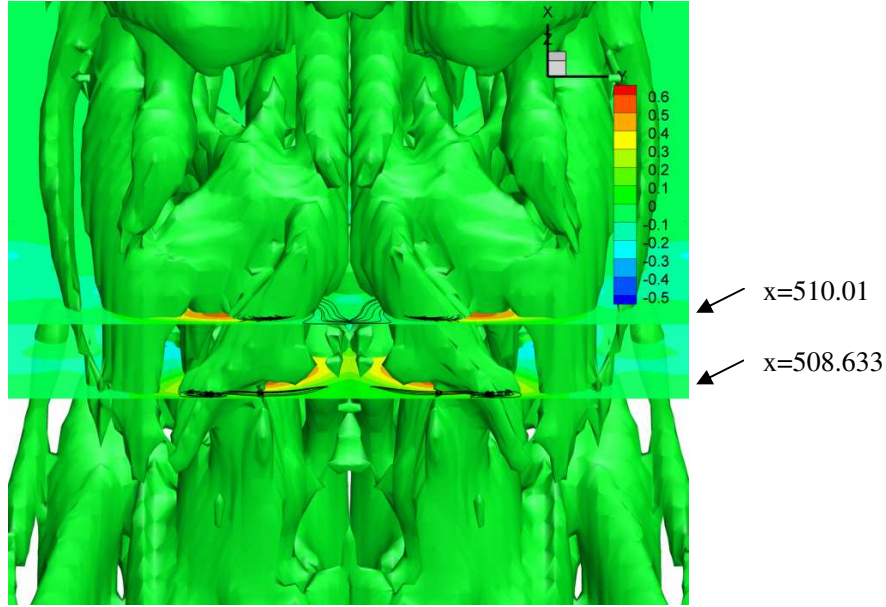

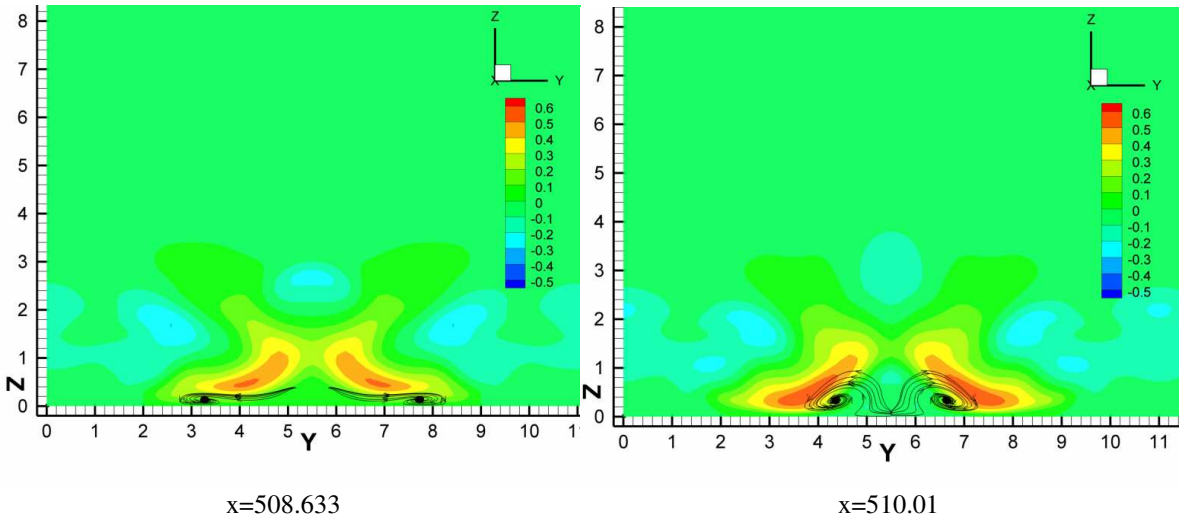

x=508.633  x=510.01
**Figure 18. Bottom view at t=8.075T and two slices of velocity distribution**

Figures 19 (a) and (b) show the streamwise velocity contours at same stremawise location x=508.663, but different time steps at t=8.0T and 8.175T. They illustrate that the high speed regions (red color) are reduced in the spanwise direction at y=4 and y=7. Figures 19 (c) and (d) show the streamwise velocity contours at same spanwise location y=4, but different time steps at t=8.0T and 8.175T. On the other hand, the HS region (red color) moves downstream (the tail moves from x=507 to x=508.5) and becomes weakened. From the above analysis based on comparison of instantaneous fields calculated for various time instants, we can find that the low-speed streaks and high-shear layers associated with the sweep 2 and ejection 2 events seem to become weaker when the ring-like vortex propagates downstream and to be withers. This can happen when the rings do not keep perfectly circular and perpendicular in standing. Due to the instability of high shear layer, the rear part (tail of HS) will eventually form small length scales at the bottom of the boundary layer.

It also gives the answer to the mystery (Guo et al, 2010) that the positive spike moves downstream rapidly at the same speed as the vortex ring (almost same as the main mean flow). This is because that the second sweep is induced by ring-like vortices, although the mean flow velocity profile is much smaller in the bottom of the boundary layer than in the upper bound of the boundary layer. Meanwhile, the shape of the positive spike looks like an olive with different sizes in the spanwise and streamwise directions, which is quite different from the shape of the original vortex legs ( $\Lambda$ shape).

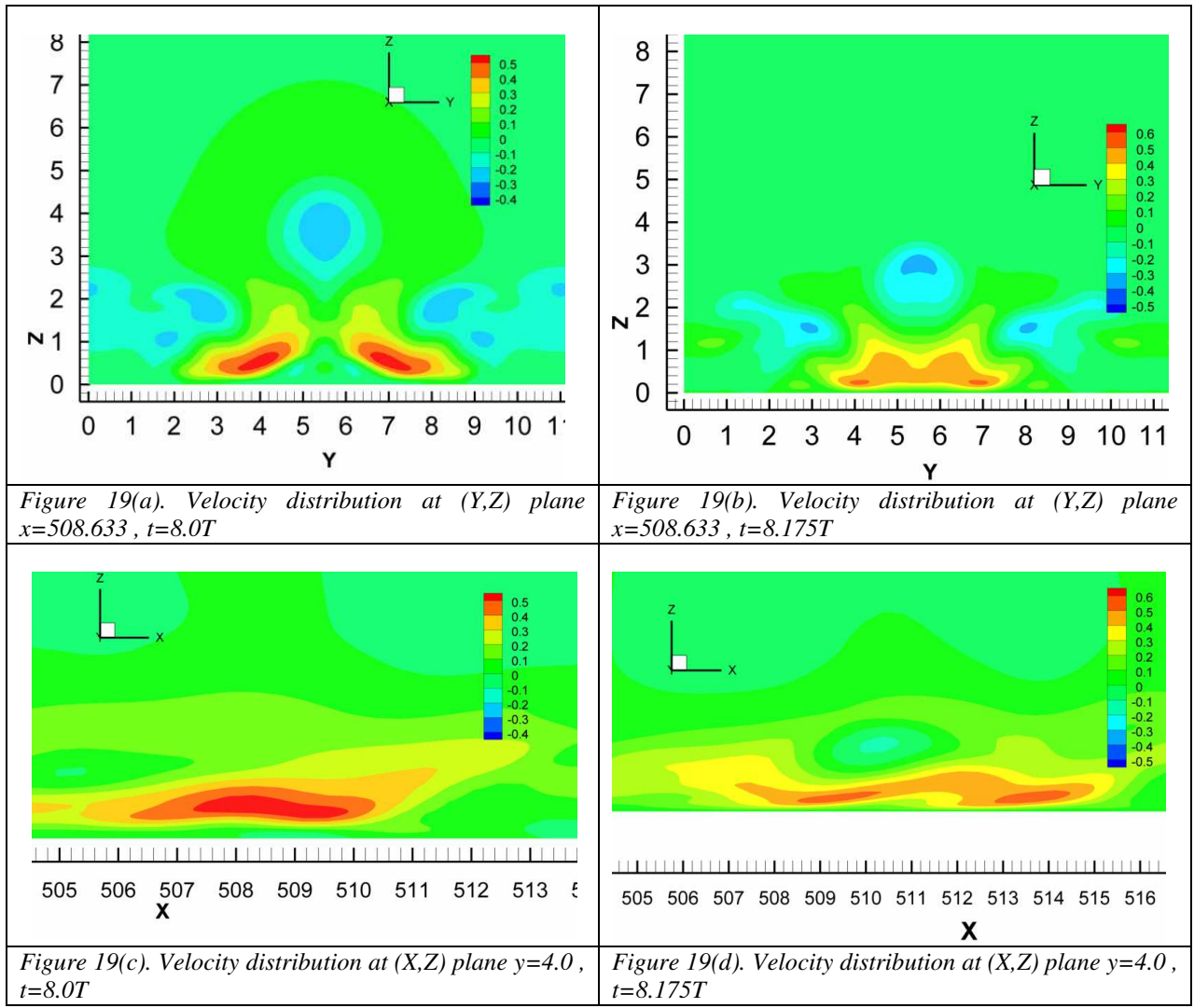

*Figure 19(a). Velocity distribution at (Y,Z) plane x=508.633 , t=8.0T*

*Figure 19(b). Velocity distribution at (Y,Z) plane x=508.633 , t=8.175T*

*Figure 19(c). Velocity distribution at (X,Z) plane y=4.0 , t=8.0T*

*Figure 19(d). Velocity distribution at (X,Z) plane y=4.0 , t=8.175T*

## V. Some visualization of small length scales for late stages at t=15.0 T

Three-dimensional high-shear layers represent the most typical indication for development of coherent structures in transitional boundary layers (similar to Rist et al, 2002). When we look at the late stage flow transition at t=15.0T (Figure 20), we can see that the high-shear layers especially the one near wall surface have broken down (HS breakdown not vortex breakdown) to form much more smaller vortices than at the time step t=8.0T .

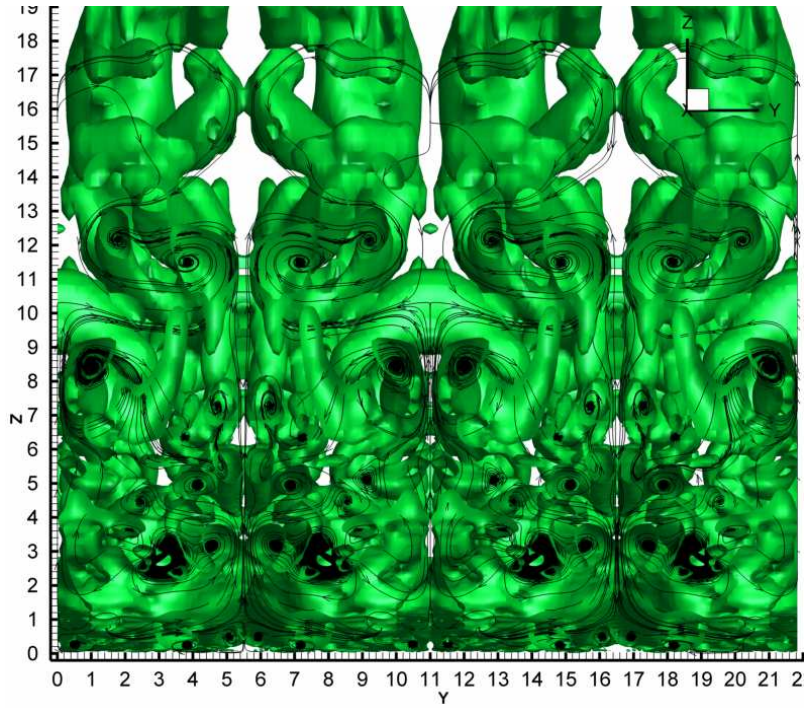

(a)

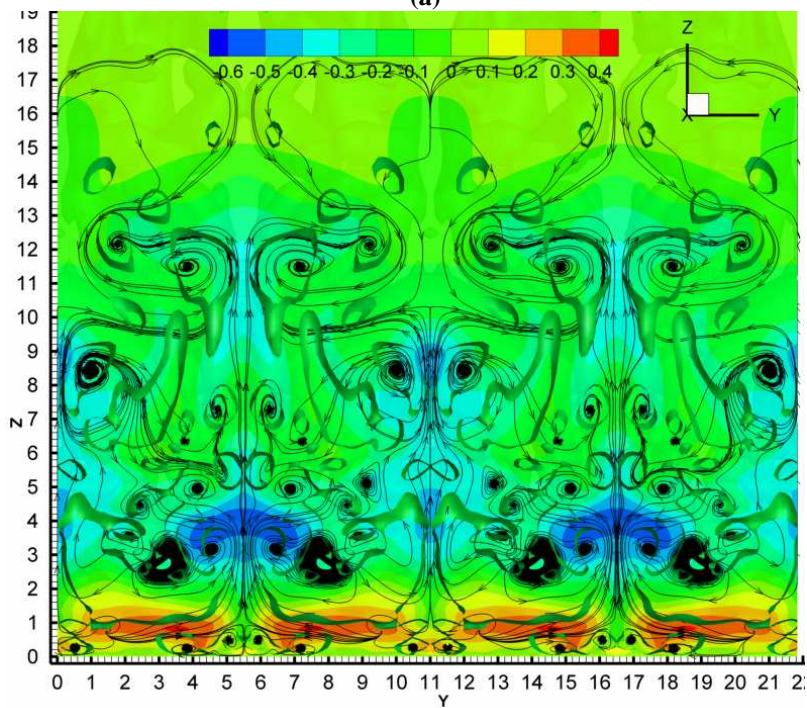

(b)

**Figure 20. Visualization of isosurface $\lambda_2$ and velocity slice at x=508.633 for (Y,Z)-Plane t=15.0T**

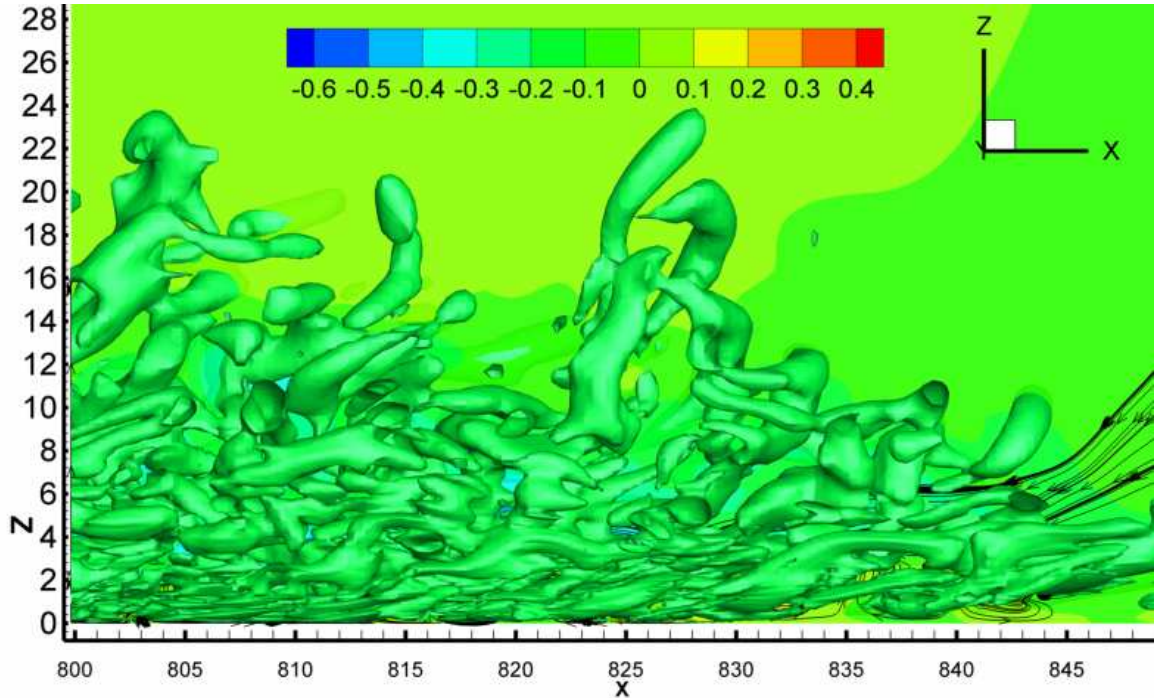

(a)

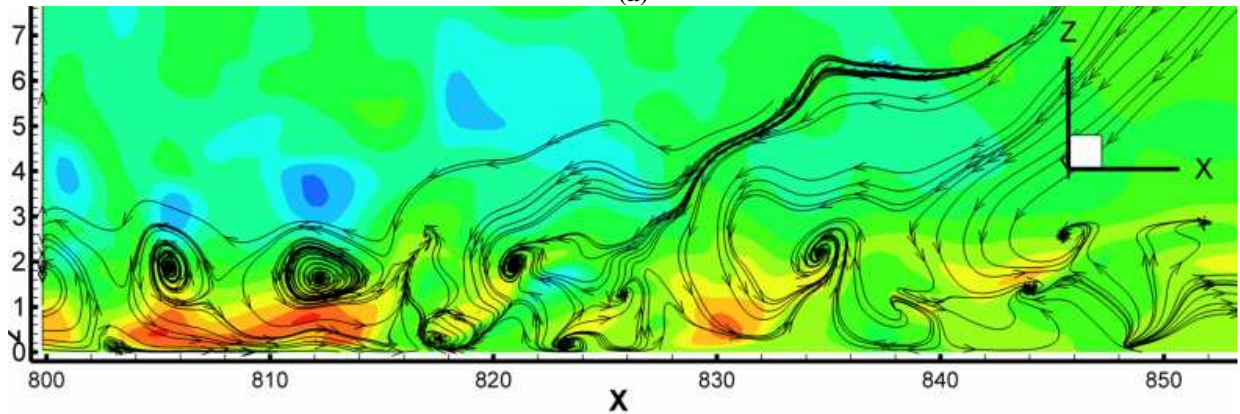

(b)

**Figure 21. Visualization of isosurface $\lambda_2$ and velocity slice at y=4.0 for (X,Z)-Plane t=15.0T**

It is emphasized that the high shear layer represents the most active and the fastest changing structures in the region near the bottom of a transitional boundary layer. All small length scale structures are generated around the HS.

Sweeping events change the shape of the velocity profile and form high shear layers near the wall due to the momentum and energy transformation. From Figures 21 depicted at the time step t=15.0T, we can observe that lots of small length scales are already generated around HS and most of them appear between the bottom of the wall and the high shear layer region and some of them are located above the HS. Since this is a 2-D visualization, the top and bottom vortex sections could belong to a same vortex.

## VI. Small vortices are generated by the wall surface at different time steps

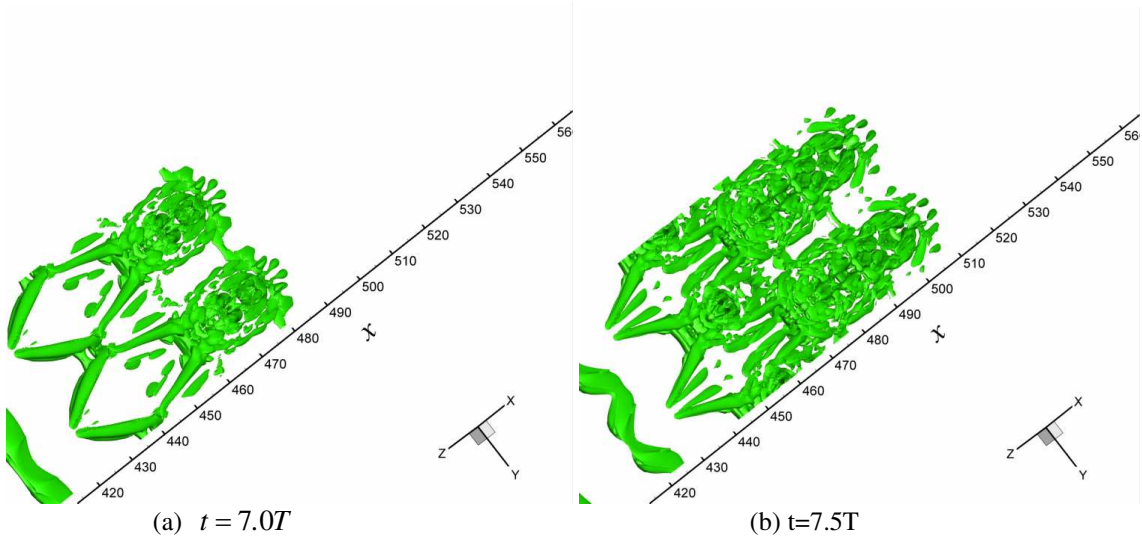

(a) $t = 7.0T$        (b) t=7.5T

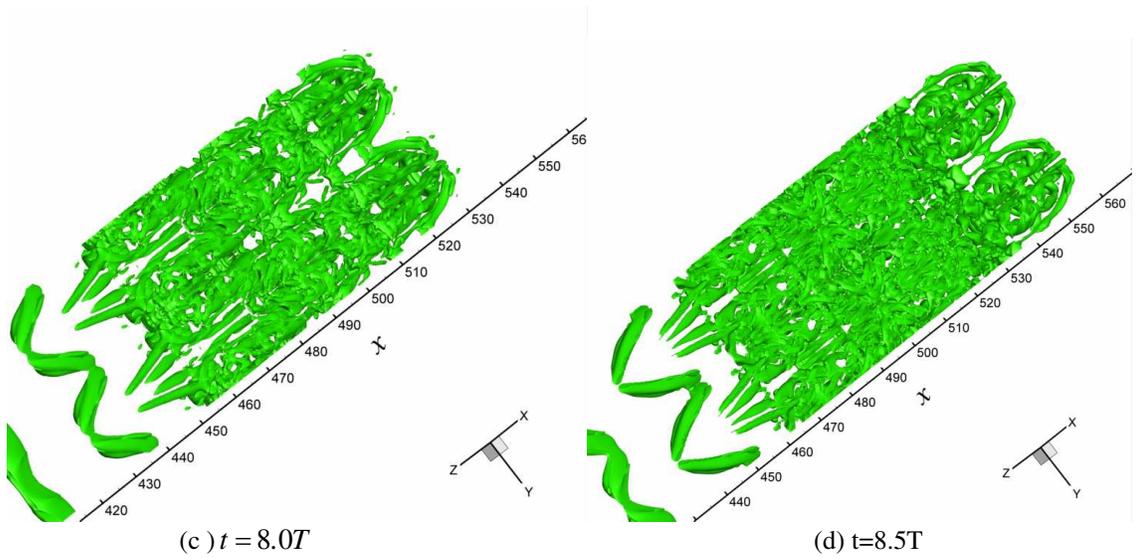

(c) $t = 8.0T$        (d) t=8.5T

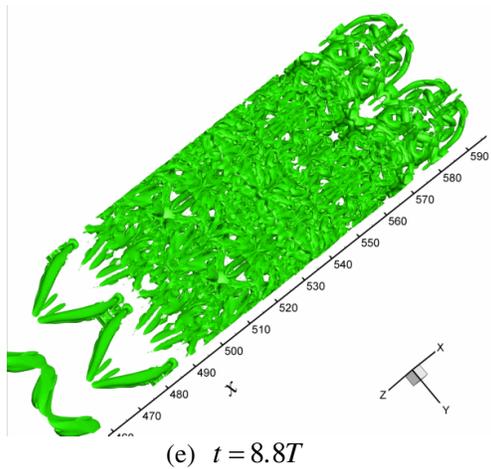

(e) $t = 8.8T$

**Figure 22. Small length scale vortex generation at different time steps (view up from bottom): small length scale vortices are generated by the solid wall near the ring necks from the beginning to the end**

All evidences provided by our new DNS confirm that the small length scale vortices are generated near the bottom (Figure 22). Thus, the main outcome of the present study is that the ring-like vortices continue playing an important role in the turbulence production mechanism occurring near the wall. The vortex rings excite the wall region strongly by downdraft jets which bring high-speed fluid to the bottom of the boundary layer, producing very intensive perturbations of velocity and high shear layer. Due to the instability of high shear layer and the interaction of HS with the wall surface, the small length scales are generated consequently.

## VII. No small length scales are produced by heading primary rings

We concentrated on examination of relationship between the downdraft motions and small length scale vortex generation and found out the physics of the following important phenomena. When the primary ring is perpendicular and perfectly circular, it will generate a strong second sweep which brings a lot of energy from the inviscid area to the bottom of the boundary layer and makes that area very active. However, when the heading primary ring is no longer perfectly circular and perpendicular but is skewed and sloped, the second sweep becomes weak simultaneously. As a result, the small length scale structures rapidly damp. Meanwhile, we clearly find the originally existing coherent structure that U-shaped vortices become clear as part of the large scale structure. This shows that the leading rings do not break down to small pieces as shown in some textbooks and research papers, but kill the small length scale as they are deformed and sloped ( Figure 23).

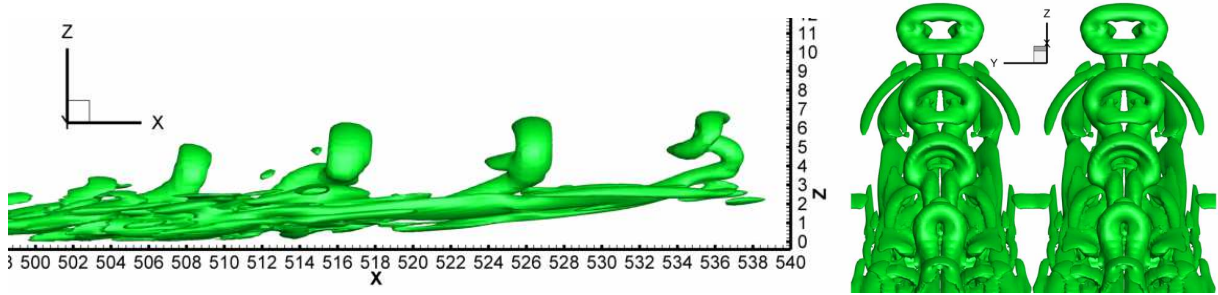

**Figure 23. Visualization of isosurface $\lambda_2$ at side view and top view t=8.0T**

## VIII. Conclusion

Despite the observation of HS in 1962, it has not attracted much research attentions for a long time either experimentally or numerically. We find that HS plays a vital role for the late stages of laminar-turbulent transition. Based on our new DNS study, the following conclusions can be made.

1. The leading rings are almost perfectly circular which are not deformed because they are located near the inviscid region which is isotropic.

2. The leading rings stand almost perpendicularly not with 45 degrees because both the top and the bottom of the ring are located in the upper bound of the boundary layer and move at a same speed (U=1).

3. Some later rings do not stand perpendicularly because they are not located fully in the inviscid region. The top of the ring moves faster than the bottom of the ring due to the boundary layer mean velocity profile. The ring legs are inclined because they are located inside the boundary layer and must be stretched and inclined.

4. The positive spikes move very fast at the same speed as the negative spikes (original vortex rings). This is because the positive spike is produced by second sweep which bring high speed from the inviscid region. There is "no shock waves" as some research papers suggested.

5. The positive spikes look like Olive but not like the original vortex legs (Λ shape) with different sizes in the span-wise direction. This is because they are secondary (not original), produced by second sweep, and the rear part (tail) will disappear due to the instability of HS.

6. The positive spikes generate high shear (HS) layers near the wall due to zero speed on the wall.

7. HS is unstable especially for those near the wall surface due to KH instability.

8. HS can breakdown and form many small vortices. Each of them has two legs which lie down on the wall surface. All small length scales are generated by HS.

9. The structure of hairpin vortex is very stable and "hairpin vortex breakdown" is never observed.

10. All small vortices are generated around the high shear layers.

11. The vortex ring must be perfectly circular and perpendicularly standing. Otherwise, if the ring is deformed and/or the standing position is inclined, the second sweep and then the positive spikes will be weakened. Then the small length scales quickly damp.

The mechanism of small length scale (turbulence) generation can be described as follows: perfect circular and perpendicular vortex rings generate strong second sweeps, then second sweeps bring the high energy from the inviscid region to the bottom of the boundary layer and generate the strong positive spikes. The positive spikes generate high shear layers which is not stable (K-H instability), so that HS and wall surface generate more and more small length scales near the wall surface. We will further study how the flow lost of the symmetric shape and become randomized.

Therefore, the mechanism of small length scale (turbulence) generation in a transitional boundary layer is revealed.

## Acknowledgments

This work was supported by AFOSR grant FA9550-08-1-0201 supervised by Dr. John Schmisseur. The authors are grateful to Texas Advanced Computing Center (TACC) for providing computation hours.